\documentclass[twocolumn,showpacs,showkeys,preprintnumbers,amsmath,amssymb]{revtex4}

\usepackage{array}
\usepackage{booktabs}
\usepackage{tabu}
\usepackage{dcolumn}
\usepackage{amsmath}
\usepackage{amsfonts}
\usepackage{amssymb}
\usepackage{graphicx}
\usepackage{subfigure}
\usepackage{graphicx}
\usepackage{dcolumn}
\usepackage{bm}

\begin{document}

\title{ Wormhole solution in modified teleparallel-Rastall gravity and energy conditions }

\author{N. Nazavari}
  \email{nazavari.ph@gmail.com}
  \author{K. Saaidi}
  \email{ksaaidi@uok.ac.ir}

  \author{A. Mohammadi}
  \email{abolhassanm@gmail.com}

\affiliation{
Department of Physics, Faculty of Science, University of Kurdistan, Sanandaj, Iran.\\}
\date{\today}

\begin{abstract}
The possibility of static and spherically symmetric traversable wormhole solution in modified teleparallel Rastall gravity (MTRG) in a non-diagonal tetrad is investigated.  Rastall assumption modifies the field equations and  energy-momentum conservation law. These modifications lead to different exact asymptotically flat traversable wormhole solutions. By imposing different constraints on modified energy-momentum conservation law, different exact solutions are found and the obtained results for teleparallel Rastall gravity (TRG), with specific choices of $f(T)$, are studied. It is shown that the Rastall parameters  have a key role in these model and in all of those exact traversable wormhole solutions the null energy condition (NEC) and weak energy condition (WEC) of matter or energy which has surrounded the wormhole are valid at the throat of the wormhole and through the space time as well.
\end{abstract}
\maketitle

\section{Introduction}
Gravity, which is one of the four basic forces in nature, is a challenging topic in physics. Therefore, many attempts have been made to understand gravity in the theory of general relativity, that was proposed by Einstein in 1916 \cite{A. E}. According to the Einstein  theory of gravity (ETG), the effect of gravity is determined by the curvature of space-time. This means that the presence of matter induces gravity by curving the space-time. \\
One of the first prediction of this theory is that, a sufficiently compact mass can deform space-time and create a black hole. One of the characteristics of ETG is called the event horizon, the boundary of the region from which no escape is possible \cite{book,Wald}.
A wormhole is known as another solution of the Einstein field equations which connects two regions of space-time by a bridge. This assumptive object connects the two points away in a universe or two different universes through a tunnel-like space \cite{Flamm, A. E2, MW}. The first introduced wormhole  was non-traversable, meaning that there can not be any signal sent through it. However, in 1988, the concept of traversable wormhole was introduced by Morris and Thorne \cite{MT,MTY}. They presented a modern wormhole solution by applying a reverse procedure technique. Instead of assuming an energy-momentum tensor and extracting the corresponding geometry, they presumed a geometry and sought to find the related stress-energy tensor.   \\
Morris and Throne took the following static, non-rotating, symmetrical metric
\begin{equation}\label{1a}
ds^2 =  e^{a(r)} \; dt^2 - e^{b(r)} \; dr^2 - r^2 \big( d\theta^2 + \sin^2(\theta) \; d\varphi^2 \big),
\end{equation}
as a geometry that fits the Einstein field equation. The radial coordinate stands in the range $r_0 < r < \infty$, where $r_0$ is the minimum of the radial coordinate which corresponds to the throat of the wormhole. The functions $a(r)$ and $b(r)$ are two arbitrary functions of the radial coordinate $r$. The former concerns with the gravitational redshift and so it is called the redshift function, and the latter is determined the shape of the wormhole throat and is addressed as the shape function.  In the literature, the shape function is usually redefined by $e^{-b(r)} = \left( 1 - {\beta(r)/ r} \right)$. This metric, Eq. (\ref{1a}), describe the traversable wormhole, if the functions $a(r)$ and $b(r)$ satisfy the following constraint:
\begin{enumerate}
\item  The event horizon should be avoided in the structure of wormholes, therefore, redshift function, $a(r)$,  should be finite at $r \geq r_{0}$. Then, existence of the traversable wormhole relies on the absence of horizon. Therefore, wormhole horizons are traversable, on both sides while the horizon of the black hole is  only one-way traversable.
\item The throat condition  implies  $\beta(r_0) = r_0$, and the shape function should satisfy the condition $\beta(r) < r$ for $r>r_0$.
\item The shape function also should follow the flare-out condition stated as $\big( \beta(r) - r \beta'(r) \big) / \beta^2(r) > 0$, where prime indicates derivative with respect to the radial coordinate $r$. At the throat, $r = r_0$, the flare-out condition is rewritten as $\beta'(r_0) < 1$.
\end{enumerate}
 As we know, the null, weak, strong and dominant energy conditions are satisfied by the ordinary (barionic matter)form of matter. In general, NEC and WEC for the ordinary matter are expressed as $T_{\mu \nu} k^\mu k^\nu \geq 0$, where $k^\mu$ is a null or time like vector respectively.
 On the other hand, based on Morris and Thorne works in the frame of ETG, to have a traversable wormhole,
   the normal energy conditions, such as  the NEC and WEC  are violated  \cite{G9, Ku10, BL11, JL12, LL13,  S14, M15, G16, R17, P18, F19, FL20}.  Such matter is addressed as "exotic matter". Their consideration came to this conclusion that the traversable wormholes are supported by the presence of exotic matters near the throat of the wormhole.\\
Although the classical forms of matters follow the energy conditions, there are cases with violation of the energy conditions in quantum field theory,  such as Casimir effect and Hawking evaporation \cite{EV, BD}. Moreover, in the field of cosmology, the existence of exotic matter sounds to be required to explain the  accelerated expansion phase. Dark energy models are usually parameterized by the parameter $\omega = p/\rho$, and $\omega < -1/3$ is required to explain the positive cosmic expansion \cite{S1, S2, LA, T26, P27, S3, E29}. Therefore, some specific types of exotic matter, such as phantom with $\omega < -1$, could explain cosmic expansion and also supports the existence of traversable wormholes
\cite{SN, ST, P30, Z31, Hawking}.
In recent years, the possibility of making wormholes  supported by ordinary matter has become an important and attractive topic in the field of physics of wormholes. Morris and Thorne tried to minimize the violation of the energy conditions by introducing different geometries. With the introduction of thin-shell wormholes, it became possible to construct a wormhole which limits the violation of energy conditions to only a thin shell \cite{MV33, MW34}. Some researchers have tried to solve the problem of exotic matter by investigating the traversable wormhole in modified theories of gravity  \cite{L35, ER36, MH37, KT38, SG39, GS40, GL41, GL42, FL43, MC44, TS45, KL46, PR47, PK48}. The field equation in the modified gravity could be rewritten in the same form as the Einstein gravity, where there is an effective stress-energy tensor $T_{\mu\nu}^{\rm eff}$, instead of the ordinary one. The effective tensor $T_{\mu\nu}^{\rm eff}$ includes two parts as: first the ordinary stress-energy tensor and the second is some modified terms with geometrical origin. The violation of NEC here is expressed in terms of the effective stress-energy tensor as $T_{\mu\nu}^{\rm eff} k^\mu k^\nu < 0$. Therefore, there may be cases where the NEC is satisfied by the matter part of the effective stress-energy tensor, although it is broken in general. During the past couple of years, people have found wormholes that satisfy the NEC in general or at a specific range \cite{RM49, ST50, K51, S52}.\\
In general relativity, a  fundamental assumption  is that the covariant divergence of the energy-momentum tensor vanishes,  i.e.,
\begin{equation}\label{1ab}
\tau ^{\nu}_{\phantom{\nu}\mu ;\nu}=0
\end{equation}
Some models beyond the standard model of gravity,  have proposed that the matter energy-momentum tensor may not covariantly conserved, such as $ f(R, T) $ gravity \cite{HLNO52}, $  f(R,L_{m}) $ gravity \cite{HL56, HL57}, and the Rastall gravity \cite{PS53}.
Rastall generalized the matter energy momentum conservation law  as follows
\begin{equation}\label{2ab}
\nabla_{\mu} \tau ^{\mu \nu} = \tilde{\lambda} \nabla ^{\nu}R,
\end{equation}
where $R$ is the Ricci scalar and $\tilde{\lambda}$ is Rastall parameter. In Rastall model of gravity because of (\ref{2ab}), the Einstein field equations are modified to the following form
\begin{equation}\label{3ab}
G_{\mu \nu}+\tilde{\lambda} \tilde{\kappa} R g_{\mu \nu} = \tilde{\kappa} T_{\mu \nu}
\end{equation}
$ \tilde{\kappa}$ is the Rastall gravitational coupling constant. \\
Since the appearance of Rastall gravity, some modifications of this theory have been proposed. For example, the authors of \cite{MH58} have generalized  the Rastall's theory by assuming $\nabla _{\mu}T^{ \mu \nu} =\nabla _{\nu}(\lambda ' R) $, where $\lambda'$ is a function of space-time coordinates. In \cite{LQ59} has been assumed,  $\nabla _{\mu}T^{ \mu \nu} =\lambda \nabla _{\nu}f(R) $, and based on this  the  electrically and magnetically neutral regular black hole solution is presented.
Also, according to  Rastall assumption, the Rastall modified version of teleparallel gravity is introduced in \cite{SN}.\\ 

  In this work we try to find a traversable  wormhole solution in modified teleparallel Rastall gravity (MTRG) for non-diagonal tetrad fields. In our point of view, the coupling between matter and curvature of space-time has an important role in gravitational phenomena and therefore in this manuscript, we consider the possibility of static and spherically symmetric traversable wormhole solutions in MTRG and we want to find different asymptotically flat geometry which describe traversable wormholes and NEC and WEC of matters that surrounded the throat of the wormhole are satisfied.\\
The paper is presented in the following order. In Section II, {a brief review on the Weitzenböck geometry and the field equations is presented. Then, we introduce the Rastall term that} appears in the conservation law equation and modifies the field equations. In Section III,  we introduce different constraint and assumptions to solve the conservation law of energy-momentum tensor and for teleparallel and modified Rastall gravity, the status of the NEC and WEC are investigated. Finally, in Section IV, conclusions are presented.


\section{Modified  Teleparallel Rastall gravity(MTRG): a review}
The defined metric $g_{\mu \nu}$ on the manifold could be expressed in terms of the tetrad fields
$e^{i}_{\phantom{i}\mu}$ and the Minkowski metric $\eta_{ij} = diag(-1, 1, 1, 1)$ as
\begin{equation}\label{2a}
g_{\mu \nu}=\eta _{i j} e^{i}_{\phantom{i}\mu}  e^{j}_{\phantom{j}\nu},
\end{equation}
where, Greek $(\mu, \nu, \ldots =0,1,2,3)$  and Latin $(i, j, \ldots =0,1,2,3)$ alphabets show space-time index and tangent space index respectively. The Weitzenböck connection, which is defined as
  \begin{equation}\label{3a}
 \Gamma^{\alpha}_{\phantom{\alpha}\mu\nu} =e _{i} ^{\phantom{i}\alpha} \partial _{\nu} e^{i} _{\phantom{i}\mu}=-e^{i}_{\phantom{i}\mu} \partial _{\nu} e _{i} ^{\phantom{i}\alpha},
\end{equation}
is the connection in the teleparallel theory, which has non-zero torsion but zero curvature. The torsion tensor is described {through the connections and} is read as
 \begin{equation}\label{4a}
 T^{\sigma} _{\phantom{\sigma}\mu \nu} \equiv \Gamma ^{\sigma}_{\phantom{\sigma} \nu \mu} -\Gamma ^{\sigma}_{\phantom{\sigma}\mu \nu} =e_{i}^{\phantom{i}\sigma} \left( \partial_{\mu} e^{i} _{\phantom{i}\nu}- \partial_{\nu} e^{i}_{\phantom{i}\mu} \right).
\end{equation}
The Weitzenbock connection is related to the Levi-Civita connection,
$  \bar{\Gamma}^{\sigma}_{\phantom{\sigma}\mu \nu}$, of General Relativity by the following relation
 \begin{equation}\label{5a}
 \bar{\Gamma}^{\sigma}_{\phantom{\sigma}\mu \nu}= \Gamma ^{\sigma}_{\phantom{\sigma}\mu \nu}- K^{\sigma}_{\phantom{\sigma}\mu \nu},
\end{equation}
where $K^{\sigma} _{\phantom{\sigma}\mu \nu}$ is known as the contorsion tensor, given by
\begin{equation}\label{6a}
K^{\sigma} _{\phantom{\sigma}\mu \nu} = \frac{1}{2} \left(T_{\mu} ^{\phantom{\mu}\sigma}{_{\nu}} +T_{\nu} ^{\phantom{\nu}\sigma}{_{\mu}} - T^{\sigma} _{\phantom{\sigma}\mu \nu}\right).
\end{equation}
The torsion scalar is expressed as
 \begin{equation}\label{7a}
T = S^{\sigma \mu \nu} T _{\sigma \mu \nu},
\end{equation}
where $ S^{\sigma \mu \nu}$ is called the super-potential and is given by
 \begin{equation}\label{8a}
 S^{\sigma \mu \nu} =-S^{\sigma \nu \mu} =\frac{1}{2} \Big( K^ {\mu \nu \sigma} -g^{\sigma \nu} T^{\alpha \mu}_{\phantom{\sigma \mu}\alpha} +g^{\sigma \mu} T^{\alpha \nu}_{\phantom {\alpha \nu} \alpha}\Big) .
\end{equation}
The action of the modified teleparallel gravity is presented by
\begin{equation}\label{9a}
S= S_{G} + S_{m} =\frac{1}{4\kappa} \int e \; f(T) d^{4} x  + \int eL_{m} d^{4} x,
\end{equation}
{in which} $f(T)$ is an arbitrary function of the torsion scalar, $e$ is the determinant of the tetrad field ${e}^{a}_{\phantom{\sigma}\mu}$ and $L_m$ is the \textbf{matter Lagrangian}. Taking variation of the action with respect to the tetrad field leads one to the following  field equation
\begin{eqnarray}\label{10a}
 S_i{}^{\mu\nu} f_{TT} \partial_\mu T + e^{-1} \partial_{\mu}(e S_i{}^{\mu\nu}) f_T
 - T^{\sigma}{}_{\mu i} S_{\sigma}{}^{\nu\mu} f_T \\
   - \frac{1}{4}e_i{}^{\nu}f= - \kappa \Theta^{\nu}_{\mu}, \nonumber
\end{eqnarray}
where $ \Theta^{\nu}_{\mu}$ is addressed to the usual energy-momentum tensor of the perfect fluid. It is shown that
\begin{eqnarray}\label{11a}
\Big(S_i{}^{\mu\nu} f_{TT} \partial_\mu T + e^{-1} \partial_{\mu}(e S_i{}^{\mu\nu}) f_T
 - T^{\sigma}{}_{\mu i} S_{\sigma}{}^{\nu\mu} f_T \\
  - \frac{1}{4}e_i{}^{\nu}f \Big)_{;\nu} =0, \nonumber
\end{eqnarray}\label{field_equation}
where the semicolon denotes the covariant derivative that in teleparallel formalism is defined as
\begin{equation}\label{12a}
V^{\mu}_{\phantom{\mu};\nu}=\partial_{\nu}V^{\mu}+ (\Gamma ^{\mu}_{\phantom{\mu}\lambda \nu}- K^{\mu}_{\phantom{\mu}\lambda \nu})V^{\lambda},
\end{equation}
for any space-time vector $V^{\mu}$. From the above result, it is realized that the covariant derivative of the energy-momentum tensor vanishes as well, i.e.
 \begin{equation}\label{13a}
\Theta ^{\nu}_{\phantom{\nu}\mu ;\nu}=0,
\end{equation}
{Then, we have the same conservation equation for energy-momentum tensor as one has in Einstein theory.} \\
The conservation equation in Einstein theory of gravity was first questioned by Rastall \cite{PS53} and he replaced it by $T^\mu_{\nu \; ; \mu} = \lambda R_{,\nu}$. This assumption, which indicates  an interaction between matter and geometry, leads to a modified field equation. Here, following Rastall assumption, we are going to presume the same case for the modified teleparallel gravity, and suggest a link between the matter and geometry  through the scalar torsion of geometry where the divergence of the energy-momentum tensor $\Theta ^{\nu}_{\phantom{\nu}\mu}$ is proportional to the divergence of the torsion scalar
\begin{equation}\label{14a}
\Theta ^{\nu}_{\phantom{\nu}\mu ;\nu}=\lambda h(T)_{,\mu}\;,
\end{equation}
where $\lambda$ is a real constant and $h(T)$ is an arbitrary analytical function of torsion.  Then, the field equation (\ref{10a}) is rewritten as
\begin{eqnarray}\label{15a}
S_i{}^{\mu\nu} f_{TT} \partial_\mu T + e^{-1} \partial_{\mu}(e S_i{}^{\mu\nu}) f_T
 - T^{\sigma}{}_{\mu i} S_{\sigma}{}^{\nu\mu} f_T \\
 -  \frac{1}{4}e_i{}^{\nu}f - \delta ^{\nu}_{\mu} \; \kappa \; \lambda \; h(T) = - \kappa \Theta^{\nu}_{\mu}. \nonumber
\end{eqnarray}
Here, $\kappa$ is the gravitational constant in Rastall theory, which in a comparison with the Newtonian gravity is expressed as
\begin{equation}\label{16a}
\kappa = {4 \gamma - 1 \over 6\gamma - 1} \; \kappa_G,
\end{equation}
where $\gamma = \lambda \kappa$ and $\kappa_G$ is the Einstein coupling constant $\kappa_G = 4 \pi G$. The energy-momentum tensor $\Theta^{\nu}_{\mu}$ is assumed to describe a nonisotropic fluid as
\begin{equation}\label{17a}
\Theta^{\nu}_{\mu}=(\rho + p_{t})u_{\mu} u^{\nu} - p_{t} \delta^{\nu} _{\mu} + (p_{r} - p_{t}) v_{\mu} v^{\nu},
\end{equation}
where $u_{\mu}$ is the time-like four velocity vector and $v_{\mu}$ stands for the unitary space-like vector in the radial direction. They satisfy the relation $u_{0}u^{0}=-v_{1} v^{1}=1$. Also, $\rho$ is the energy density, and $p_{r}$ and $p_{t}$ indicate the radial and transverse pressure respectively. Using Eqs. (\ref{14a}) and (\ref{17a}), one can obtain the conservation relation for energy-momentum tensor as
\begin{equation}\label{18a}
\bar{p}_r' + \frac{a'}{2 } (\bar{\rho} + \bar{p}_r) -\frac{2}{r}(\bar{p}_t-\bar{p}_r)  = 0 ,
\end{equation}
here $\bar{\rho}= \rho - \lambda h(T)$ and $\bar{p}_{r.t}= p_{r,t} +\lambda h(T)$. It is clearly seen that for $\lambda =0$ this relation is reduced to ordinary form of energy-momentum conservation relation. \\
A common way to describe a traversable wormhole is through a static, spherically symmetric metric expressed by Eq. (\ref{1a}). There is no unique set of tetrad fields for the metric, and we only assume a non-diagonal tetrad field as
\begin{equation} \label{19a}
{e}^{a}_{\phantom{\sigma}\mu}=
\left(\begin{array}{cccc}
e^{a/2} & 0 & 0&0\\
0& e^{b/2}\sin\theta \cos\varphi & r \cos\theta \cos\varphi & -r \sin\theta \sin \varphi \\
0 &  e^{b/2} \sin \theta \sin \varphi & r \cos \theta \sin \varphi & r \sin\theta \cos \varphi \\
0 & e^{b/2} \cos \theta &-r \sin \theta &0\\
\end{array}\right).
\end{equation}
in which $g_{00}=e^{a(r)}$ and  $g_{11}= e^{b(r)}$. The determinant of the above tetrad field is given by $e=det[{e}^{a}_{\phantom{\sigma}\mu}]= r^{2}\sin\theta e^{(a+b)/2}$. The torsion scalar is obtained from Eqs. (\ref{4a}), (\ref{7a}), and (\ref{8a}) as
\begin{equation}\label{20a}
 T(r)=\frac{2e^{-b}}{r^{2}}\Big(e^{\frac{b}{2}} -1\Big)\Big(e^{\frac{b}{2}} -1-ra'\Big).
\end{equation}
Substituting Eqs. (\ref{17a}), (\ref{19a}), and (\ref{20a}) in the field equation (\ref{2a}), the nonzero components of the energy-momentum tensor are obtained as
\begin{eqnarray}
\kappa \rho(r)&=& \frac{e^{-b/2}}{r}(1-e^{-b/2}) f_{T}'  -\left(\frac{T}{4}-\frac{1} {2r^2}\right)f_T  \label{21a}\\
   & &  +\frac{e^{-b}}{2 r^2} \left(rb '-1\right)f_T + \frac{f}{4} + \gamma h(T),  \nonumber \\
\kappa  p_r(r) &=& \left[\frac{T}{4}- \frac{1}{2r^2}+\frac{e^{-b}}{2r^2}
  (1+ra')\right]f_T-\frac{f}{4} - \gamma h(T), \label{22a} \\
\kappa  p_t(r) &=& \frac{e^{-b}}{2}\left(\frac{a'}{2}+\frac{1}{r}-\frac{e^{b/2}}{r}\right)f_{T}' -  \frac{f}{4} - \gamma h(T)  \label{23a} \\
   & +&  f_T\left\{ \frac{T}{4}+\frac{e^{-b}}{2 r} \left[\left(\frac{1}
 {2}+\frac{ra'}{4}\right) \left(a'-b'\right)+\frac{ra''}{2}\right]\right\}, \nonumber
\end{eqnarray}
which in general are expressed in terms of the redshift and shape functions. Also, the Rastall term $\gamma h(T)$ and the coefficient $\kappa$, which appears in equations, will affect the behavior and even the magnitude of the components. These terms might also change the behavior of the energy conditions. So in order to find a \textbf{traversable wormhole} solutions we solve Eq. (\ref{18a}) with various assumptions for different types of $f(T)$ function .
\section{Wormhole solution in MTRG}
In this section, we make different suggestions and solve modified conservation law of matter energy-momentum tensor, Eq.(\ref{18a}). {Then, the obtained results for different constraint are} investigated for different types of $f(T)$ function.
\subsection{$\lambda T'\;h_T(T) =0$. }
 By replacing Eqs. (\ref{21a}), (\ref{22a}) and (\ref{23a}) into Eq. (\ref{18a}) one can find
\begin{equation}\label{24a}
\lambda T' h_T(T) = 0.
\end{equation}
For $\lambda \neq 0$, the solution of Eq. (\ref{24a}) is as $T'=0$ or $h_{T}(T)=0$. {However,} $h_{T}(T)=0$ is not a suitable solution, because in this case the Rastall assumption is removed. Therefore the solution of Eq. (\ref{24a}) is $T'=0$ only. This solution shows us, $T(r)= T_{0}$ and for simplicity we assume $T_{0}=0$; this means $T=0$. As usual, the static and spherically symmetric wormhole is given by Eq. (\ref{1a}) and by  the following definition for $e^{b(r)}$,
\begin{equation}\label{25a}
e^{b(r)}= \frac{1}{1- \frac{\beta(r)}{r}}.
\end{equation}
Imposing $T'=T=0$ and Eq. (\ref{25a}), we rewrite the energy-momentum tensor components, Eqs. (\ref{21a})-(\ref{23a}) as
\begin{eqnarray}
\kappa  \rho(r) & = & \frac{\beta'}{2 r^{2}}f_{T}(0)+ \frac{f(0)}{4}+ \lambda h(0), \label{26a} \\
\kappa  p_r(r) & = &  \frac{1}{2 r^{2}}\Big[ \big(1-\frac{\beta}{r}\big)\big(1+ r a' \big)-1 \Big]f_{T}(0)-  \label{27a} \\
& &  \Big(\frac{f(0)}{4}+ \lambda h(0) \Big),\label{27a} \nonumber \\
\kappa  p_t(r) & = & \frac{1}{4r^{2}}\Big(1- \frac{\beta}{r}\Big)\Big[r^{2}a''+ \Big(1+ \frac{a' r}{2}\Big)\Big(r a'  \\
   & & + \frac{\beta- r \beta '}{(r- \beta)} \Big)\Big]f_{T}(0)- \Big(\frac{f(0)}{4}+ \lambda h(0) \Big). \nonumber \label{28a}
\end{eqnarray}
In asymptotically flat space-time, the components of energy-momentum tensor {should vanish in infinity. This feature is satisfied in Eqs. (\ref{26a})-(\ref{28a}) when} $f(0)/4+ \lambda h(0)=0$. Then, the WEC, $(\rho(r) \geq 0 $  and $ (\rho +p_{i})\geq 0 (i=r, t))$ is given by
\begin{eqnarray}
 \rho(r) & = & \frac{f_{T}(0)}{2\kappa r^{2}} \beta '\; , \label{29a} \\
\rho(r)+  p_r(r) & = & \frac{f_{T}(0)}{2 \kappa r} \Big[\frac{\beta ' r -\beta}{r}+ \big(1- \frac{\beta}{r}\big) a' \Big]\label{30a} \; ,\\
\rho(r) +  p_t(r) & = & \frac{f_{T}(0)}{2 \kappa r^{2}}\bigg\{ 1+ \frac{\beta 'r -\beta}{2r}(1- \frac{r a'}{2})-\\
 & & (1- \frac{\beta}{r})\Big[1- \frac{1}{2}\big(1+ \frac{r a'}{2}\big)r a' - \frac{r^{2}a''}{2}\Big] \bigg\} \; . \nonumber \label{31a}
\end{eqnarray}
 It is known that for a traversable wormhole $\beta(r_{0})= r_{0}$ and Eqs. {(\ref{29a})-(\ref{31a})} at the throat  of wormhole is obtained  as
\begin{eqnarray}
 \rho(r_{0}) & = & \frac{f_{T}(0)}{2 \kappa r_{0}^{2}}\beta'(r_{0}), \label{32a} \\
\rho(r_{0}) +  p_r(r_{0}) & = &  \frac{f_{T}(0)}{2\kappa r_{0}^{2}}\Big(\beta '(r_{0})-1\Big), \label{33a}\\
\rho(r_{0}) +  p_t(r_{0}) & = &  \frac{f_{T}(0)}{2 \kappa r_{0}^{2}} \Big[1- \frac{1- \beta'(r_{0}))}{2} \label{34a} \\
& & \Big(1- \frac{r_{0}\; a'(r_{0})}{2}\Big)\Big].\nonumber
\end{eqnarray}
Based on the flaring out condition at the throat, namely, $\beta '(r_{0}) <1$, in order to have $(\rho(r_{0})+ p_{r}(r_{0})) > 0$ one of the set conditions ($\kappa >0$  and $ f_{T}(0) <0$) or ($\kappa <0 $  and $  f_{T}(0) >0$) {should be imposed. On the other hand, applying these two conditions on (\ref{32a}) and using the positiveness of $\rho$, the constraint on the derivative of the form function would be tighter as $\beta'(r_{0}) < 0$.} Moreover, for $0 < \beta'(r_{0}) <1$ the quantity $f_{T}(0)/ \kappa$ should be positive and this is satisfied for ($\kappa$ and $f_{T}(0) >0$) or ($\kappa$ and $f_{T}(0) <0$). Furthermore, for having $(\rho(r_{0})+ p_{t}(r_{0}))  > 0$ beside $\beta '(r_{0}) < 0 $ and $f_{T}(0)/\kappa < 0$, we should have another relation as
\begin{equation}\label{35a}
a' \; r_{0}  < 2- \frac{4}{1- \beta '(r_{0})}.
\end{equation}
For checking the correctness of Eq. (\ref{35a}), we focus on $T=0$. We assume $e^{b/2} \neq 1$ and therefore for vanishing $T$, the following relation should be satisfied
\begin{equation}\label{36a}
e^{b/2} -1- r a'=0.
\end{equation}
As an answer, one can consider a specific form for the shape function $\beta(r)$ which is studied in \cite{BHL, Ellis}, as
\begin{equation}\label{37a}
\beta(r)= \frac{r_{0}^{2}}{r}.
\end{equation}
It is seen that $\beta(r_{0})=r_{0}$ and $\beta '(r_{0})=-1<0$, then this form of shape function satisfies the traversable condition on $\beta(r)$. Based on this, Eq. (\ref{36a}) is rewritten as
\begin{equation}\label{38a}
\frac{1}{\sqrt{1- \frac{r_0^2}{r^2}}}-1=r\; a',
\end{equation}
and by solving this equation, one can find the redshift function as
\begin{equation}\label{39a}
a(r)=\ln \Big(1+ \sqrt{1- \frac{r_0^2}{r^2}}\Big) +a_{0}.
\end{equation}
where $a_{0}$ is a constant and it is seen that the $g_{00}$ component is asymptotically flat. Inserting Eqs. (\ref{37a}) and (\ref{39a}) into Eqs. (\ref{26a})-(\ref{28a}), one can find
\begin{eqnarray}
  \rho(r) & = &  \frac{r_{0}^{2}}{2 r^{4}}\mid \frac{f_{T}(0)}{\kappa}\mid , \label{40a} \\
      p_r (r)& = &  \frac{1 }{2 r^{2}}\mid \frac{f_{T}(0)}{\kappa}\mid\Big[1-\sqrt{1- \frac{r_{0}^{2}}{r^{2}}} \Big],\label{41a} \\
   p_t (r) & = & \frac{1}{4 r^{2}}\mid \frac{f_{T}(0)}{\kappa}\mid \Big[\sqrt{1-\frac{r_{0}^{2}}{r^{2}}}-\frac{r_{0}^{2}}{r^{2}}-1 \Big].\label{42a}
\end{eqnarray}
It is well seen that for any type of  ($f_{T}(0) >0$, $\kappa < 0$) or ($f_{T}(0) <0$, $\kappa > 0$) constraints, the quantity of energy density is positive and other constraints of WEC are given by
\begin{eqnarray}
  \rho(r)+ p_r (r) & = & \frac{\mid f_{T}(0)/\kappa \mid}{2r^{2}}\Big[\frac{r_{0}^{2}}{r^{2}}- \sqrt{1- \frac{r_{0}^{2}}{r^{2}}}  +1\Big], \label{43a}\\
  \rho(r)+ p_t (r)  & = &  \frac{\mid f_{T}(0)/\kappa \mid}{4r^{2}}\Big[\sqrt{1- \frac{r_{0}^{2}}{r^{2}}}+\frac{3r_{0}^{2}}{2r^{2}}-1\Big]. \label{44a}
\end{eqnarray}
It is obviously seen that at the throat of the wormhole Eqs. (\ref{40a}), (\ref{43a}) and (\ref{44a}) are positive and proportional {to the factor} $\mid f_{T}(0)/\kappa r_0^2 \mid $, therefore NEC and WEC at the throat of wormhole and neighborhood of it are valid.
\begin{figure}[t]
\subfigure[]{\includegraphics[width=7cm]{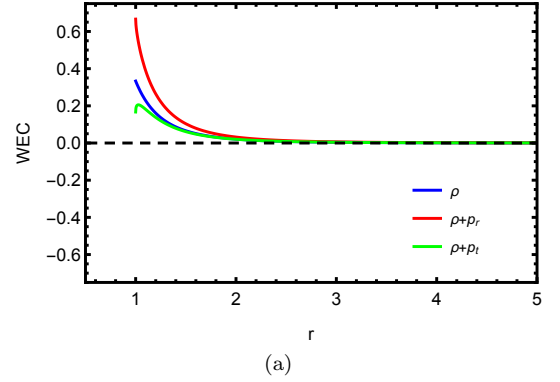}}
\caption{ WEC equation versus radial coordinates for throat radius $r_{0}=1$ with free parameter: $\kappa= -1.5$ }\label{NECA1}
\end{figure}
 We will investigate  the obtained results of \textbf{pervious section} for different form  of $f(T)$ .
\subsubsection{  $f(T)=T$}
As we know, in the standard model of gravity the traversable wormhole  is obtained only for the case that  NEC and WEC are violated. On the other hand, it is shown that  ordinary teleparallel gravity is equivalent of standard model of  general relativity \cite{BLT}. Therefore, the NEC and WEC of matter, which has surrounded the wormhole, are violated as well \cite{BHL}. Then, in this part we explore the  teleparallel Rastall gravity for obtained results of previous section.\\
The energy-momentum for the case of $T'=T=0$, in the teleparallel-Rastall  model of gravity, is given by
\begin{eqnarray}
  \rho(r) & = & \frac{\beta '}{2 \kappa r^{2}}, \label{45a} \\
 p_r (r) & = & \frac{1}{2 \kappa r^{2}}\bigg[ \Big(1- \frac{\beta}{r}\Big)\Big(1+ r a' \Big)-1 \bigg], \label{46a} \\
 p_t (r)  & = &  \frac{1}{4 \kappa r^{2}} \Big(1- \frac{\beta}{r}\Big)\bigg[r^{2}a'' + \Big(1+ \frac{r a'}{2}\Big) \label{47a} \\
& & \Big(r a'+ \frac{\beta - r \beta'}{ r- \beta}\Big) \bigg]. \nonumber
\end{eqnarray}
For having positive energy density through the space-time, we should have $\beta' /\kappa > 0$. This is satisfied for two different conditions  ($\beta' > 0$  and $ \kappa > 0$) or ($\beta' < 0$  and $ \kappa < 0$). The components of  NEC is given by:
\begin{eqnarray}
  \rho(r)+ p_r (r) & = & \frac{1}{2 \kappa r^{2}}\Big(1- \frac{\beta}{r}\Big) \Big(\frac{r \beta ' -\beta}{r- \beta}+r a'\Big),\label{48a} \\
 \rho(r) + p_t (r) & = & \frac{1}{4\kappa r^{2}}\Big(1-\frac{\beta}{r} \Big)\Big[r^{2} a''\\
 &+ & \Big(1+ \frac{ra'}{2}\Big)\Big(ra'+ \frac{\beta- r \beta '}{r-\beta}\Big)+ \frac{2r \beta '}{r- \beta}\Big].\nonumber \label{49a}
\end{eqnarray}
 Using Eqs. (\ref{37a}) and (\ref{39a}), the NEC are rewritten as:
\begin{eqnarray}
  \rho(r)+ p_r (r) & = & \frac{\mid 1/ \kappa\mid}{2  r^{2}}\Big(1-\sqrt{1- \frac{r_{0}^{2}}{r^{2}}}-\frac{r_{0}^{2}}{r^{2}} \Big),\label{50a} \\
 \rho(r) + p_t (r) & = & \frac{\mid 1/ \kappa\mid}{4 r^{2}}\Big( \sqrt{1- \frac{r_{0}^{2}}{r^{2}}}+ \frac{3r_{0}^{2}}{2 r^{2}} -1\Big).\label{51a}
\end{eqnarray}
WEC for $r_0 =1 $  and $\kappa = -1.5$  versus radial coordinate, $r$, is plotted in Fig.\ref{NECA1}.,  and it is seen that WEC and NEC of the  matter which has surrounded the wormhole in the teleparallel-Rastall model of gravity is not violated. \\

\subsubsection{ $f(T)=e^{- T/T_1}$}

We consider the shape function and redshift function given by Eqs. (\ref{37a}) and (\ref{39a}) in the modified teleparallel-Rastall gravity which $f(T)$ is given by
$$ f(T)=e^{-T/T_1},$$ {where $T_1$ is a real constant.}
\begin{figure}[t]
\subfigure[]{\includegraphics[width=7cm]{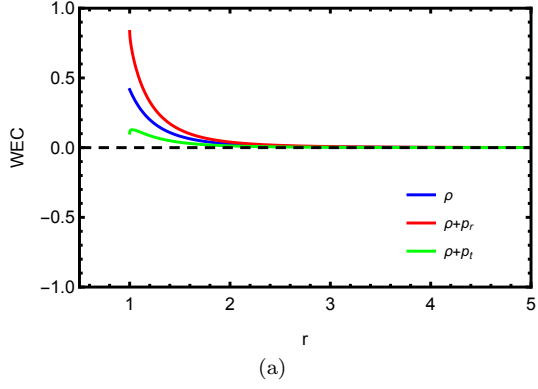}}
\caption{ WEC equation versus radial coordinates for throat radius $r_{0}=1$ with free parameter: $\kappa= \pm 1.5$, $T_{1}=\pm 0.8$ }\label{WEC2}
\end{figure}
We will investigate the obtained results of previous section for different forms  of $f(T)$.

{By substituting this function} into Eqs. (\ref{40a})-(\ref{42a}), we obtain the following expressions
\begin{eqnarray}
  \rho(r) & = &  \frac{r_{0}^{2}}{2 \kappa T_{1}\; r^{4}}, \label{52a} \\
  p_r (r)& = &  \frac{1}{2 \kappa T_{1}r^{2}}\Big(1- \sqrt{1- \frac{r_{0}^{2}}{r^{2}}} \Big),\label{53a} \\
  p_t (r) & = & \frac{1}{4\kappa T_{1}\;r^{2}}\Big(  \sqrt{1-\frac{r_{0}^{2}}{r^{2}}}-\frac{r_{0}^{2}}{2r^{2}}-1 \Big).\label{54a}
\end{eqnarray}
It can be seen from Eq. (\ref{52a}) that the value of energy density with  ($ \kappa >0$, $T_{1}>0$) or ($ \kappa <0$, $T_{1}<0$) conditions  is always positive. Using the above field equations, the NEC takes the following form,
\begin{eqnarray}
  \rho(r)+ p_r (r) & = &  \frac{ 1}{2 \kappa T_{1 } r^{2}}\Big(1-\sqrt{1-\frac{r_{0}^{2}}{r^{2}}}+ \frac{r_{0}^{2}}{r^{2}}\Big), \label{55a} \\
  \rho(r)+ p_t (r) & = &\frac{ 1}{4 \kappa T_{1 } r^{2}}\Big(\sqrt{1-\frac{r_{0}^{2}}{r^{2}}}+\frac{3r_{0}^{2}}{2r^{2}}-1\Big).\label{56a}
\end{eqnarray}
One can  obviously see that at the throat of the wormhole, Eqs. (\ref{52a}), (\ref{55a}) and (\ref{56a}) are positive and proportional to $1/\kappa T_1r_0^2 $. Therefore, NEC and WEC at the throat of wormhole and neighborhood of it for this type of modify teleparallel Rastall gravity are valid. These equations are plotted in Fig.\ref{WEC2} for $r_0 =1$ and $\kappa = \pm 1.5$ and $T_1 = \pm 0.8$.


\subsection{Constant radial pressure: $p_r(r) = p_0$}

To obtain an alternative solution for energy-momentum tensor, we assume {a constant radial pressure, i.e. $p_r(r) = p_0$ and $p_r'(r) =0$.} Then, from  Eq. (\ref{22a}), one can find the Rastall term as
\begin{equation}\label{57a}
\gamma h(T)=\Big[\frac{T}{4}- \frac{1}{2r^{2}}+ \frac{e^{-b}}{2 r^{2}}(1+ r a')\Big]f_{T}- \frac{f}{4}- \kappa p_{0}.
\end{equation}
{Also, the conservation of energy-momentum tensor is rewritten as}
\begin{equation}\label{58a}
\frac{a'}{2}\Big( \rho(r) + p_{r}(r)\Big) - \frac{2}{r}\Big( p_{t}(r)- p_{r}(r)\Big) + \gamma h'(T)=0.
\end{equation}
By doing the derivative of $\gamma h(T)$  and inserting it into Eq. (\ref{58a}), it is seen that the modified conservation energy-momentum tensor is satisfied. Therefore, by substituting Eq. (\ref{57a}) into Eqs. (\ref{21a})-(\ref{23a}) and using $e^{-b}=(1- \beta(r)/r)$, one arrives at
\begin{eqnarray}
\kappa  \rho(r) & = & \frac{1}{r}\bigg[ \sqrt{1- \frac{\beta}{r}}- \Big(1- \frac{\beta}{r}\Big) \bigg]f'_{T}- \kappa p_{0} \label{59a}\\
&+ & \frac{1}{2r}\bigg[\Big(1- \frac{\beta}{r}\Big)a'- \frac{\beta- r\beta '}{r^{2}}\bigg]f_{T}, \nonumber \\
 \kappa p_r (r) & = & \kappa p_{0}, \label{60a} \\
 \kappa p_t (r)  & = & \frac{1}{2r}\bigg[\big(1+ \frac{r a'}{2}\big)\big(1-\frac{\beta}{r}\big) - \sqrt{1- \frac{\beta}{r}}\bigg]f'_{T} \label{61a}\\
 &+ & f_{T}\bigg[\frac{1}{2r^{2}}- \frac{1}{2r^{2}}\big(1- \frac{\beta}{r}\big)\big(1+ra'\big)  \nonumber \\
   &+ &  \frac{1}{4r}\big(1-\frac{\beta}{r}\big)\big(1+ \frac{ra'}{2}\big)a'+ \frac{\beta - r \beta '}{4r^{3}}\big(1+ \frac{r a'}{2}\big) \nonumber \\
  & +&  \frac{a''}{4}\big(1- \frac{\beta}{r}\big) \bigg]+ \kappa p_{0}. \nonumber
\end{eqnarray}
Evaluating Eqs. (\ref{59a})-(\ref{61a}) at the throat of the wormhole results in the following relations
\begin{eqnarray}
 \rho(r) & = & -\frac{f_{T}}{2\kappa r_{0}^{2}}\Big(1- \beta'(r_{0}) \Big)- p_{0},\label{62a} \\
 p_r (r) & = &  p_{0}, \label{63a} \\
 p_t (r)  & = &  \frac{f_{T}}{2\kappa r_{0}^{2}}\bigg[1+ \frac{1}{2}\Big(1- \beta '(r_{0})\Big) \Big(1+ \frac{r_{0} a'(r_{0})}{2}\Big) \bigg] +p_{0}. \nonumber \\
 \label{64a}
\end{eqnarray}
For obtaining Eqs. (\ref{62a})-(\ref{64a}), the traversable wormhole condition of form function at the throat of the wormhole, i.e., $\beta(r_{0})=r_{0}$, is used. {Then, the WEC at the throat are read as}
\begin{eqnarray}
 \rho(r_{0}) & = & \frac{\mid f_{T}/\kappa\mid}{2 r_{0}^{2}}\Big(1- \beta'(r_0) \Big)- p_{0},\label{65a} \\
 \rho(r_{0})+ p_r (r_{0}) & = & \frac{\mid f_{T}/\kappa\mid}{2 r_{0}^{2}}\Big(1- \beta'(r_{0})\Big), \label{66a} \\
\rho(r_{0})+ p_t (r_{0})  & = & \frac{\mid f_{T}/\kappa\mid}{2 r_{0}^{2}}\bigg[ \frac{1}{2}\Big( \beta '(r_{0})-1\Big) \\
& &\Big(1 +\frac{1}{2}r_{0}a'(r_{0}\Big)- \beta'(r_{0})\bigg]. \nonumber \label{67a}
\end{eqnarray}
{First, we assume that $p_0$ is negative, i.e. $p_{0}<0$. So,} based on the flaring out constraint at the throat of the wormhole, i.e., $(\beta - r\beta')/2\beta ^{2} \vert_{r=r_{0}}> 0$, $\beta'(r_{0})$ should be smaller than one, $\beta'(r_{0})< 1$. Therefore, to have $\rho(r_{0})>0$, the constraint $f_{T}/ \kappa <0$ {should be imposed. On the other hand, for the case of positive pressure $p_{0}>0$, besides all the above mentioned conditions, an extra condition as $\Big(\mid f_{T}/\kappa \mid (1- \beta '(r_{0}))\Big)/2r_0^{2}  > p_{0}$ should also be applied.} \\
{From the positiveness of $\rho$, it is realized that the energy condition $(\rho(r_{0})+p_{r}(r_{0}))> 0$ is satisfied. However, an extra constraint is required to verify the other energy condition $\rho(r_{0})+p_{t}(r_{0})> 0$, which is}
\begin{equation}\label{68a}
r_{0} a'(r_{0})< \frac{2+2 \beta '(r_{0})}{\beta'(r_{0})-1}.
\end{equation}
The constant redshift function is a solution of Eq. (\ref{68a}), then  $a=constant$ impose $\beta'(r_{0})<-1$ which is satisfied by $\beta(r)=r_{0}^{2}/r$. Then in this model, the constant redshift function and $\beta(r)= r_{0}^{2}/r$ are a set of functions where the NEC and WEC are valid in the throat and its neighborhood. \\
\begin{figure}[t]
\subfigure[]{\includegraphics[width=7cm]{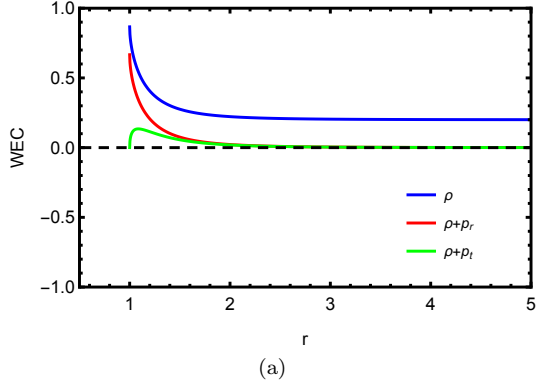}}
\caption{ WEC equation versus radial coordinates for throat radius $r_{0}=1$ with free parameter: $\kappa= - 1.5$, $p_{0}=-0.2$ }\label{WEC3}
\end{figure}
\subsubsection{$f(T)=T$.}
In this stage, we will check our obtained results in teleparallel-Rastall model of gravity for  $a=constant$ and  $\beta'(r_{0})<-1$. Inserting these functions into the equation of the energy-momentum tensor, Eqs. (\ref{59a})-(\ref{61a}), one can find the following expressions
\begin{eqnarray}
 \rho(r) & = &-\frac{r_{0}^{2}}{\kappa r^{4}}- p_{0},\label{69a} \\
 p_t (r) & = &  \frac{r_{0}^{^{2}}}{\kappa r^{4}}+p_{0}. \label{70a}
\end{eqnarray}
The WEC are given by the positiveness of Eq. (\ref{69a}) and
\begin{eqnarray}
\rho(r) +  p_r (r) =  \frac{r_{0}^{2}}{\mid\kappa\mid r^{4}}, \label{71a} \\
\rho(r) +  p_t (r)  =   0 . \label{72a}
\end{eqnarray}
Eqs. (\ref{69a}), (\ref{71a}) and (\ref{72a}),  show that  for $\kappa <0$ the NEC and WEC are satisfied throughout space-time.

\subsubsection{$f(T)= e^{-T/T_{1}}$ }
 We want to check our previous subsection's  study for $f(T)= e^{-T/T_{1}}$, $a=a_{0}$ and $\beta(r)=r_{0}^{2}/r$ functions. By substituting these functions into Eqs. (\ref{59a})- (\ref{61a}) we find the following relations
\begin{eqnarray}
 \rho(r) & = &\dfrac{1}{\kappa  T_1^2\; r}\bigg[\Big(\sqrt{1- \frac{r_0^2}{r^2}}- 1+ \dfrac{r_{0}^{2}}{r^{2}}\Big)T' \label{73a} \\
 &+& \dfrac{r_{0}^{2}T_1 }{r^{3}}\bigg]e^{-T/T_{1}}- p_{0}, \nonumber \\
 p_r(r) & = & p_{0}, \label{74a} \\
 p_t (r) & = &\dfrac{1}{2 \kappa T_1^2\; r}\bigg[\Big(1- \dfrac{r_0^2}{r^2}-\sqrt{1- \frac{r_0^2}{r^2}}\Big)T' \label{75a} \\
 &-& \dfrac{2r_0^2T_1}{r^{3}}\bigg]e^{-T/T_{1}} + p_{0} , \nonumber
\end{eqnarray}
where
\begin{eqnarray}
T &=&\frac{2}{r^{2}} \Big( \sqrt{1- \frac{r_{0}^{2}}{r^{2}}}-1 \Big)^{2}, \nonumber \\
T'& =&\frac{4r_{0}^{2}}{r^{5}}\bigg[1- \Big(\sqrt{1- \frac{r_{0}^{2}}{r^{2}}}\Big)^{-1} \bigg]- \frac{4}{r^{3}}\Big(\sqrt{1- \frac{r_{0}^{2}}{r^{2}}} -1\Big)^{2} . \nonumber
\end{eqnarray}
Also, the NEC are given by
\begin{eqnarray}
\rho(r) +p_r (r) & = &\dfrac{1}{\kappa  T_1^2\; r}\bigg[\Big(\sqrt{1- \frac{r_0^2}{r^2}}- 1+ \dfrac{r_{0}^{2}}{r^{2}}\Big)T' \label{76a} \\
 &+& \dfrac{r_{0}^{2}T_1 }{r^{3}}\bigg]e^{-T/T_{1}}, \nonumber \\
  \rho(r) +p_t (r) & = & \dfrac{1}{2\kappa T_1^2 r}\Big(1-\sqrt{1- \frac{r_{0}^{2}}{r^{2}}}\Big) \times \label{77a}\\
 & &\Big(\sqrt{1- \frac{r_{0}^{2}}{r^{2}}}\Big)T'e^{-T/T_{1}}. \nonumber
\end{eqnarray}
NEC and WEC at the throat of the wormhole are given by
\begin{eqnarray}
\rho(r_0) & = &\dfrac{1}{\kappa  T_1^2\; r_0^4}e^{-2/T_1r_0^2}\Big(T_{0}r_{0}^{2}-4\Big) - p_{0}, \nonumber \\
 \rho(r_0) +p_r(r_0) & =  &\dfrac{1}{\kappa  T_1^2\; r_0^4}e^{-2/T_1r_0^2}\Big(T_{0}r_{0}^{2}-4\Big), \nonumber \\
 \rho(r_0) +p_t (r_0) & = & - \dfrac{2}{\kappa  T_1^2\; r_0^4}e^{-2/T_1r_0^2}, \nonumber
\end{eqnarray}
It is seen that in this case, the NEC and WEC are valid for some set of conditions.


\subsection{ $p_r(r) = p_0$ and $\gamma  h(T)+ \frac{f(T)}{4}+ \kappa p_{0}=\frac{T}{2}f_{T}(T)$ }
In order to have  different solutions for our model, we impose an extra condition as
\begin{equation}\label{78a}
\gamma  h(T)+ \frac{f(T)}{4}+ \kappa p_{0}=\frac{T}{2}f_{T}(T).
\end{equation}
In this case, Eq. (\ref{57a}) is reduced to
\begin{equation}\label{79a}
-\frac{T(r)}{4}- \frac{1}{2r^{2}}+ \frac{e^{-b}}{2r^{2}}(1+ ra')=0.
\end{equation}
By using Eq. (\ref{20a}) and solving Eq. (\ref{79a}), one can find
\begin{equation}\label{80a}
r a'=\frac{2\Big(1-e^{-b/2}\Big)}{ e^{-b/2}},
\end{equation}
{Then, from Eq. (\ref{80a}) and for}  $e^{-b}=(1-\beta/r) = (1-r_0^2/r^2)$, we obtain
\begin{equation}\label{82a}
e^{a(r)}=\; \Big(1-\sqrt{1- \frac{r_{0}^{2}}{r^{2}}}\Big)^{2},
\end{equation}
where $e^{a(r)} \rightarrow 1$ as $r \rightarrow \infty$. \\
{Now, we are going to} solve Eqs. (\ref{21a})-(\ref{23a}) for the typical functions of $f(T)$, form function $\beta(r) = r_0^2/r$ and Eq. (\ref{82a}) as redshift function.
\subsubsection{$f(T) = T$}
  As a specific case, we assume $f(T)=T$ then according to Eq. (\ref{78a}) one can find $\gamma h(T)=(T/4- \kappa p_{0})$ and Eqs. (\ref{59a})-(\ref{61a}) are rewritten as
\begin{eqnarray}
 \rho(r)  & = & \frac{\mid 1/ \kappa \mid}{r^{2}} \Big(1- \sqrt{1- \frac{r_{0}^{2}}{r^{2}}}\Big)- p_{0}, \label{83a}\\
 p_r (r) & = & p_{0}, \label{84a}\\
p_t (r) & = &\frac{\mid 1/ \kappa \mid}{r^{2}}\Big[ 2 \Big(\sqrt{1- \frac{r_{0}^{2}}{r^{2}}}-1\Big)+ \frac{r_{0}^{2}}{r^{2}}\Big]+p_{0},\label{85a}
\end{eqnarray}
the components of NEC are given by the followings relations
\begin{eqnarray}
 \rho(r)  +p_r(r) & = &  \frac{\mid 1/ \kappa \mid}{r^{2}} \Big(1- \sqrt{1- \frac{r_{0}^{2}}{r^{2}}}\Big),\label{86a}\\
\rho(r) + p_t(r)  & = & \frac{\mid 1/ \kappa \mid}{r^{2}}\Big( \sqrt{1- \frac{r_{0}^{2}}{r^{2}}}+ \frac{r_{0}^{2}}{r^{2}}-1\Big). \label{87a}
\end{eqnarray}
From Eqs. (\ref{86a}), (\ref{87a}), and (\ref{83a}), it is transparent that NEC and WEC are satisfied throughout the space-time {by applying} the constraints $\kappa < 0$ and $p_0 < 2/\mid\kappa \mid r_0^2$. This fact is shown in Figure \ref{WEC3}.\\
{At the throat of the wormhole, the NEC are}
\begin{eqnarray}
 \rho(r)  & = & \frac{\mid 1/ \kappa \mid}{r_0^{2}} - p_{0}, \nonumber \\
 \rho(r)  +p_r(r) & = &  \frac{\mid 1/ \kappa \mid}{r_0^{2}}, \nonumber \\
\rho(r) + p_t(r)  & = & 0,  \nonumber
\end{eqnarray}
where the positiveness of them is clearly seen.
\subsubsection{$f(T) = e^{-T/T_1}$}
In this stage, we proceed in an attempt to analytically solve the field Eqs. (\ref{21a})-(\ref{23a}) by considering a function $f(T)=e^{-T/T_1}$. According to Eq. (\ref{78a}), we have $\gamma h(T)=-e^{-T/T_1}/2 \;(2\; + T/T_{1}) -\kappa p_{0}$.  Therefore,  Eqs.(\ref{59a})-(\ref{61a}) are rewritten as
\begin{eqnarray}
 \rho(r) & = & \frac{ e^{-T/T_1}}{\kappa\; T_1^2\;r^{2}}\; \Big[ r \Big(\frac{r_{0}^{2}}{r^{2}}+ \sqrt{1-\frac{r_0^2}{r^2}}-1\Big)T' \label{88a} \nonumber \\
 &+ &\Big(1- \sqrt{1-\frac{r_0^2}{r^2}}\Big)\; T_{1} \Big]- p_{0}, \label{88a} \\
p_r(r)  & = & p_{0}, \label{89a} \\
 p_t(r)  & = & \frac{ e^{-T/T_1}}{\kappa\; T_1^2\;r^{2}}\Big[2\Big(\sqrt{1-\frac{r_0^2}{r^2}}-1 \Big)+ \frac{r_{0}^{2}}{r^{2}} \Big]+p_{0}. \label{90a}
\end{eqnarray}
where
\begin{eqnarray}
 T(r) & = &  -\frac{2}{r^{2}}\; \Big(\sqrt{1- \frac{r_0^2}{r^2}} -1 \Big)^2, \nonumber \\
T'(r)  & = & \frac{4}{r^{3}}\Big(\sqrt{1- \frac{r_0^2}{r^2}} -1 \Big)^{2}+ \frac{4r_{0}^{2}}{r^{5}}\Big[ \Big(\sqrt{1- \frac{r_0^2}{r^2}}\Big)^{-1}-1\Big]. \nonumber
\end{eqnarray}

In this case, the components of NEC {are read as}
\begin{eqnarray}
\rho(r)+ p_r(r)  & = & \frac{ e^{-T/T_1}}{\kappa\; T_1^2\;r^{2}}\; \Big[ r \Big(\frac{r_{0}^{2}}{r^{2}}+ \sqrt{1-\frac{r_0^2}{r^2}}-1\Big)T' \nonumber \\
 &+ &\Big(1- \sqrt{1-\frac{r_0^2}{r^2}}\Big)\; T_{1} \Big], \label{93a} \\
\rho(r)+p_t(r)  & = & \frac{ e^{-T/T_1}}{\kappa\; T_1^2\;r^{2}} \Big[\Big(\frac{r_{0}^{2}}{r^{2}}+ \sqrt{1-\frac{r_0^2}{r^2}} -1 \Big) \nonumber \\
& \times & \Big(T_{0}+ r\; T'\Big)\Big]. \label{94a}
\end{eqnarray}

The WEC at the throat takes the form:
\begin{eqnarray}
 \rho(r_{0}) & = & \frac{1}{ \kappa \;T_{1}^{2}\;r_{0}^{4}}e^{2/(r_{0}^{2}T_{1})}\Big(4+ r_{0}^{2}T_{0}\Big)- p_{0},\label{95a}\\
 \rho(r_{0}) +p_r (r_{0})& = & \frac{1}{ \kappa \;T_{1}^{2}\;r_{0}^{4}}e^{2/(r_{0}^{2}T_{1})}\Big(4+ r_{0}^{2}T_{1}\Big),\label{96a}\\
  \rho(r_{0})+p_t(r_{0}) & = &   \frac{1}{ \kappa \;T_{1}^{2}\;r_{0}^{4}}e^{2/(r_{0}^{2}T_{1})}. \label{97a}
\end{eqnarray}
Note that for $p_0 < 0$ that the energy density  at the throat of wormhole and neighborhood is positive  by imposing specific sets of free parameters. Therefore, it is seen that  NEC and WEC in this case are valid.

\section{Conclusion }
During this work, the energy conditions for the wormhole solutions in the frame of modified teleparallel Rastall gravity have been studied. The main dynamical variable in the teleparallel gravity are the tetrad field which are built based on the metric. However, there is no unique tetrad field for the presumed metric and one could have both diagonal or non-diagonal tetrad field. \\
{A well-known static spherically symmetric metric was utilized to establish a non-diagonal tetrads. Then, the fields equation were written based on this non-diagonal tetrads.} Also, it was assumed that the conservation of the energy-momentum  tensor is no longer valid and the covariant derivative of the energy- momentum  tensor is modified by adding a Rastall term to the equation. {The Rastall term includes a constant $\lambda$, which can take both positive an negative values, and an arbitrary function of torsion scalar $h(T)$.}  Due to this Rastall term, the field  equations and the Newtonian gravitational constant have been  modified as well. {Considering the modified conservation and field equations, we studied the wormhole solutions for three different constraints. Different geometries were found that can describe traversable wormholes.} Using the obtained geometries, we found the components of NEC and WEC in general form. \\
{The obtained wormhole solution for a static and spherical symmetric geometry} in standard model of Einstein theory of gravity and teleparallel gravity, which is equivalent with Einstein theory, indicates that the NEC and WEC of matter, that surrounds the throat of  wormhole, are violated. Therefore we tried to test our obtained results for {teleparallel gravity, $f(T)=T$, and a specific function of modified $f(T)$ gravity as $f(T) = e^{(-T/T_1)}$.}

The first constraint which is obtained from conservation law of energy-momentum tensor is $\lambda T'h_T(T) =0$. By applying this constraint on field equation and assuming specific form function, we obtained an asymptotically flat geometry  which describe a traversable wormhole and in contrast to ordinary teleparallel gravity, in teleparallel-Rastall gravity (TRG) and MTRG the NEC and WEC are not  violated.

The second assumption that we imposed on conservation relation or energy-momentum tensor, was $p_r = p_0$. For this suggestion, we found an asymptotically flat geometry wormhole solution which the traversable constraints are satisfied. Moreover for a specific set of free parameters of the TRG and MTRG models the NEC and WEC of the matter or energy which has surrounded the throat of  wormhole are valid. Finally by introducing the Rastall term such as
$$\gamma h(T) = Tf_T(T)/2- f(T)/4-\kappa p_0$$
we explored an alternative static spherical symmetric   geometry  which is asymptotically flat and   a traversable wormhole is described. In addition, NEC and WEC are satisfied for a specific set of parameters.

\section{Declarations }
The authors declare that there is no conflict of interests regarding the publication of this paper.


\end{document}